\documentstyle[11pt,amssymb]{article}

\let\a=\alpha    
    
  \let\n=\nu

\def\nn{\nonumber} \def\bd{\begin{document}} \def\ed{\end{document}}
\def\ds{\documentstyle} \let\fr=\frac \let\bl=\bigl \let\br=\bigr
\let\Br=\Bigr \let\Bl=\Bigl 
\let\bm=\bibitem
\let\na=\nabla
\let\pa=\partial \let\ov=\overline 
\newcommand{\be}{\begin{equation}} 
\newcommand{\ee}{\end{equation}} 
\def\ba{\begin{array}}
\def\ea{\end{array}}
\def\ft#1#2{{\textstyle{{\scriptstyle #1}\over {\scriptstyle #2}}}}
\def\fft#1#2{{#1 \over #2}}
\def\del{\partial}
\def\vp{\varphi}
\def\sst#1{{\scriptscriptstyle #1}}
\def\oneone{\rlap 1\mkern4mu{\rm l}}
\def\td{\tilde}
\def\wtd{\widetilde}
\def\ie{\rm i.e.\ }
\def\dalemb#1#2{{\vbox{\hrule height .#2pt
        \hbox{\vrule width.#2pt height#1pt \kern#1pt
                \vrule width.#2pt}
        \hrule height.#2pt}}}
\def\square{\mathord{\dalemb{6.8}{7}\hbox{\hskip1pt}}}
\newcommand{\ho}[1]{$\, ^{#1}$}
\newcommand{\hoch}[1]{$\, ^{#1}$}
\newcommand{\bea}{\begin{eqnarray}} 
\newcommand{\eea}{\end{eqnarray}} 
\newcommand{\ra}{\rightarrow}
\newcommand{\lra}{\longrightarrow}
\newcommand{\Lra}{\Leftrightarrow}
\newcommand{\ap}{\alpha^\prime}
\newcommand{\bp}{\tilde \beta^\prime}
\newcommand{\tr}{{\rm tr} }
\newcommand{\Tr}{{\rm Tr} } 
\def\0{{\sst{(0)}}}
\def\1{{\sst{(1)}}}
\def\2{{\sst{(2)}}}
\def\3{{\sst{(3)}}}
\def\4{{\sst{(4)}}}
\def\5{{\sst{(5)}}}
\def\6{{\sst{(6)}}}
\def\7{{\sst{(7)}}}
\def\8{{\sst{(8)}}}
\def\n{{\sst{(n)}}}
\def\tV{\widetilde V}
\def\tW{\widetilde W}
\def\tH{\widetilde H}
\def\tE{\widetilde E}
\def\tF{\widetilde F}
\def\tA{\widetilde A}
\def\im{{{\rm i}}}
\def\tY{{{\wtd Y}}}
\def\ep{{\epsilon}}
\def\vep{{\varepsilon}}
\def\R{\rlap{\rm I}\mkern3mu{\rm R}}
\def\cD{{\cal D}}
\def\semi{{\ltimes}}

\newcommand{\NP}{Nucl. Phys. }
\newcommand{\tamphys}{\it Center for Theoretical Physics,
Texas A\&M University, College Station, TX 77843}
\newcommand{\upenn}{\it Dept. of Phys. and Astro.,
University of Pennsylvania,
Philadelphia, PA 19104}

\newcommand{\auth}{M. Cveti\v{c}\hoch{\dagger1},
H. L\"u\hoch{\dagger1} and C.N. Pope\hoch{\ddagger2}}

\begin{document}
\begin{flushright}
\hfill{CTP TAMU-10/00 \\
UPR/881-T \\
March 2000}\\
\hfill{\bf hep-th/0003286}\\
\end{flushright}

\vspace{10pt}

\begin{center}
{\large {\bf Consistent Kaluza-Klein Sphere Reductions}}

\vspace{20pt}

\auth

\vspace{10pt}
{\hoch{\dagger}\upenn}

\vspace{10pt}
{\hoch{\ddagger}\tamphys}

\vspace{30pt}

\underline{ABSTRACT}
\end{center}

     We study the circumstances under which a Kaluza-Klein reduction
on an $n$-sphere, with a massless truncation that includes all the
Yang-Mills fields of $SO(n+1)$, can be consistent at the full
non-linear level.  We take as the starting point a theory comprising a
$p$-form field strength and (possibly) a dilaton, coupled to gravity
in the higher dimension $D$.  We show that aside from the
previously-studied cases with $(D,p)=(11,4)$ and $(10,5)$ (associated
with the $S^4$ and $S^7$ reductions of $D=11$ supergravity, and the
$S^5$ reduction of type IIB supergravity), the only other
possibilities that allow consistent reductions are for $p=2$, reduced
on $S^2$, and for $p=3$, reduced on $S^3$ or $S^{D-3}$.  We construct
the fully non-linear Kaluza-Klein Ans\"atze in all these cases.  In
particular, we obtain $D=3$, $N=8$, $SO(8)$ and $D=7$, $N=2$, $SO(4)$
gauged supergravities from $S^7$ and $S^3$ reductions of $N=1$
supergravity in $D=10$.

{\vfill\leftline{}\vfill
\vskip 10pt \footnoterule {\footnotesize \hoch{1} Research supported
in part by DOE grant DOE-FG02-95ER40893
\vskip  -12pt} \vskip   14pt
{\footnotesize
        \hoch{2}        Research supported in part by DOE
grant DOE-FG03-95ER40917 \vskip -12pt}  \vskip  14pt
}

\pagebreak
\setcounter{page}{1}

\section{Introduction}

    Much progress has been achieved recently in understanding the full
non-linear structure of certain Kaluza-Klein sphere reductions.  To be
specific, we have in mind the remarkable cases where it is consistent
to include lower-dimensional fields in the reduction Ansatz that
parameterise inhomogeneous deformations of the internal sphere metric.
Generically, one would expect that performing a truncation of the
complete Kaluza-Klein towers of massless and massive modes to the
purely massless sector would give rise to inconsistencies beyond the
linearised level, since currents built from the massless fields would
act as sources for the massive fields that have been set to zero.
Indeed this is exactly what usually happens; one cannot make a
consistent Kaluza-Klein $n$-sphere reduction of a generic theory in which
all the massless fields, including, in particular, the full set of
$SO(n+1)$ gauge fields are retained.  However, in certain very
special cases a consistent reduction of this kind is possible.

   An important early example of this type was much studied in the
1980's, namely the seven-sphere compactification of eleven-dimensional
supergravity.  It was first shown at the level of linearised
fluctuations \cite{dupo} around the AdS$_4\times S^7$ Freund-Rubin
\cite{frru} vacuum solution that the massless modes described
four-dimensional $N=8$ gauged $SO(8)$ supergravity \cite{dwn2}.
Subsequently, it was shown that despite all the apparent obstacles,
the reduction to the massless $N=8$ multiplet can be carried through
as an exact embedding at the full non-linear level \cite{dwn},
although the construction is an extremely complex one.  It has long
been believed that consistent reductions should also be possible in
the case of the $S^5$ compactification of type IIB supergravity, and
the $S^4$ compactification of eleven-dimensional supergravity, to
yield the maximal gauged $SO(6)$ supergravity in $D=5$, and the
maximal gauged $SO(5)$ supergravity in $D=7$, respectively.  Indeed,
the consistent $S^4$ reduction Ansatz from $D=11$ has recently been
constructed \cite{vann1,vann2}. No analogous construction exists for the
complete massless reduction on $S^5$.

      It is sometimes helpful to study sphere reductions in which only
a subset of the complete set of massless fields is retained, in such a
way that one still has the non-triviality of the inhomogeneous sphere
deformations, while at the same time making the problem of obtaining
an explicit reduction Ansatz more tractable and manageable.  This can
be especially useful if one actually wants to use the Ansatz for the
purpose of lifting solutions of the lower-dimensional theory back to
the higher dimension, in which case full explicit reduction formulae
are highly advantageous.  In this spirit, consistent reductions in the
three cases mentioned above were constructed for truncations in which
only the maximal abelian subgroups $U(1)^4$, $U(1)^3$ and $U(1)^2$ of
the full $SO(8)$, $SO(6)$ and $SO(5)$ gauge groups were retained,
together with associated scalar fields \cite{ten}.  The $U(1)^3$
example provided the first concrete evidence for the consistency of
the $S^5$ reduction of type IIB supergravity.  The reduction Ans\"atze
were sufficiently explicit that they could be used for the purpose of
lifting certain AdS black-hole solutions back to the higher dimension
\cite{ten}, where they become rotating D3-branes
\cite{rd31,rd32,cg,ten} and M-branes \cite{rms,ten}.

    Other consistent reductions involving subsets of the complete
massless sector have subsequently been constructed, including an $S^4$
reduction to give $N=2$ gauged $SU(2)$ supergravity in $D=7$
\cite{d7gauge}; an $S^5$ reduction to $N=4$ gauged $SU(2)\times U(1)$
supergravity in $D=5$ \cite{d5gauge}; and an $S^7$ reduction to $N=4$
gauged $SO(4)$ supergravity in $D=4$ \cite{d4gauge}.  In addition, the
$N=2$ gauged $SU(2)$ supergravity in $D=6$ was obtained {\it via} a
consistent reduction from massive type IIA supergravity on a locally
$S^4$ internal space \cite{d6gauge}.  This is actually the largest
possible supersymmetry for a gauged theory in $D=6$, even though the
maximum supersymmetry for ungauged supergravity is $N=4$.

    Somtimes, It can also be useful to construct a Kaluza-Klein sphere
reduction in which a non-supersymmetric truncation of the massless
supermultiplet is made.  One example of this type involved truncating
the maximal supergravities in $D=4$, 5 and 7 to a subsector comprising
just gravity and a set of 7, 5 or 4 scalars respectively.  These scalars
correspond to the diagonal subset of fields in the unimodular 
symmetric tensors $T_{ij}$ describing the scalars in the
$SL(8,\R)/SO(8)$, $SL(6,\R)/SO(6)$ and $SL(5,\R)/SO(5)$ scalar
submanifolds of the full supergravities.  In \cite{cglp,clps}, the
full non-linear reduction Ans\"atze for these embedding were
constructed, and proved to be consistent.  

   Another example of a non-supersymmetric truncation was constructed
in \cite{clpst}, where the full set of twenty scalars $T_{ij}$ of the
coset $SL(6,\R)/SO(6)$ were retained in an $S^5$ reduction from
$D=10$.  Consistency now requires that one include also the full set
of $SO(6)$ Yang-Mills gauge fields.  In fact only the metric and
self-dual 5-form of the type IIB supergravity are involved in this
reduction, and it can equivalently be viewed as a Kaluza-Klein reduction
of a theory of pure gravity plus self-dual 5-form in $D=10$, with all
massless fields retained in $D=5$.  (The truncation of type IIB
supergravity to just gravity and the self-dual 5-form is itself a
consistent one in $D=10$.)  The self-duality of the 5-form is crucial
for the consistency of the reduction.

    One should not conclude from the listing of examples above that
consistent Kaluza-Klein sphere reductions are a commonplace.  In fact,
if we restrict attention to cases where one starts in the higher
dimension with just gravity and a $p$-form field strength, then it
turns out that the only cases that can give consistent reductions are
related to the examples mentioned above.\footnote{To be precise, we should
emphasise that what we are discussing here is cases where {\it all} of
the $SO(n+1)$ Yang-Mills gauge fields associated with the isometry
group of the $n$-sphere are retained in the truncation, together with
other associated massless scalars.}  The reason for this can be
understood as follows.  For reductions of the type we are considering,
where the lower-dimensional theory obtained by the $S^n$ reduction
has an $SO(n+1)$ local gauge symmetry, it is essential that the {\it
ungauged} theory that would result from performing a reduction on the
$n$-torus rather than the $n$-sphere should have scalars described by
a coset manifold $G/H$ such that $H$ at least contains $SO(n+1)$.  The
reason for this is that in the process of gauging the ungauged theory,
a subgroup $SO(n+1)$ of the global symmetry $G$ must become local, and
this subgroup must be contained within $H$.  Now if a generic theory
of gravity and antisymmetric tensors is reduced on $T^n$, it will give
rise to a lower-dimensional theory with a $GL(n,\R)$ global symmetry
\cite{cremmer,julia,cjlp1}, for which the maximal compact subgroup is
$SO(n)$.  This is insufficient for allowing an $SO(n+1)$ gauging.
Note that in particular this argument shows that it is not possible
to perform a consistent $n$-sphere reduction of a pure gravity theory,
in which the Yang-Mills fields of $SO(n+1)$ are retained.

   In certain very special theories, the $GL(n,\R)$ global symmetry
arising from a $T^n$ reduction is enhanced to a larger symmetry, as a
result of ``conspiracies'' between scalars coming from the
Kaluza-Klein reduction of the metric and of the other
higher-dimensional fields.  However, as discussed in
\cite{cjlp1,cjlp3}, such cases are very few and far between.  In
particular, if we consider a $D$-dimensional theory consisting of
gravity and a single $p$-form field strength, an enhancement of the
global symmetry can occur {\it only} if $(D,p)$ is equal to $(11,4)$,
$(11,7)$ or $(10,5)$.  Since a 7-form in $D=11$ is dual to a 4-form
this means that the only cases with symmetry enhancements are
associated with $D=11$ supergravity and type IIB supergravity.  The
corresponding enhanced symmetries in each case are to $SL(8,\R)/SO(8)$
in $D=4$, $SL(5,\R)/SO(5)$ in $D=7$, and $SL(6,\R)/SO(6)$ in
$D=5$.\footnote{Actually further conditions must be fulfilled in order
for the symmetry enhancements to take place.  In the case of $D=11$
reduced to $D=4$ on $T^7$, the enhancement of $GL(7,\R)$ to $SL(8,\R)$
requires that the 4-form have an $FFA$ term in $D=11$ with precisely
the coefficient dictated by supersymmetry; this in fact means that the
enhanced symmetry is even larger, namely $E_7$.  For the $D=10$ theory
reduced on $T^5$, the enhancement from $GL(5,\R)$ to $SL(6,\R)$
requires that the 5-form be self-dual (or anti-self-dual).}  These
enhancements then allow the $SO(8)$, $SO(6)$ and $SO(5)$ gaugings,
respectively.\footnote{The importance of enhancements of the global
symmetry in toroidal reductions was also observed in \cite{nasvam},
although it was assumed there that the phenomenon was much more
widespread than is actually the case.}

       Note that whereas there is a consistent $S^5$ reduction of
gravity with a self-dual 5-form in which {\it only} the metric, the
$SO(6)$ Yang-Mills fields $A_\1^{ij}$ and the 20 scalars $T_{ij}$ are
retained, the situation is a little different in the $S^4$ and $S^7$
reductions from $D=11$.  In addition to keeping the corresponding
Yang-Mills fields $A_\1^{ij}$ and scalars $T_{ij}$, the consistency of
the $S^4$ reduction requires also keeping the five 3-forms $A_\3^i$ of
the seven-dimensional theory, whilst the $S^7$ reduction instead
requires also keeping the 35 pseudoscalars $\phi_{[ijk\ell]_+}$
(self-dual in the $SO(8)$ indices).  These additional fields are
needed in the reductions because the Yang-Mills fields act as sources
for them \cite{clpst}.  In fact we can summarise the situation in all
three of these examples of the $S^4$, $S^7$ and $S^5$ reductions as
follows.  In all cases, the consistent $n$-sphere reduction that includes
all the Yang-Mills fields of $SO(n+1)$ requires one to include {\it
all} the massless fields in the lower-dimensional theory.
Thus in the $S^5$ case, if we reduce the theory
of gravity and the self-dual 5-form then the metric, the $SO(6)$ Yang-Mills
fields and the scalars $T_{ij}$ indeed constitute the complete set of
massless fields in five dimensions.  In the $S^4$ reduction from $D=11$
the five 3-forms $A_\3^i$ are massless too, and indeed they must be
included also in the consistent reduction.  Likewise, in the $S^7$
reduction from $D=11$ the 35 pseudoscalars $\phi_{[ijk\ell]_+}$ are
also massless, and indeed they must be included in the consistent reduction.

    Further possibilities for consistent $n$-sphere reductions in
which all the $SO(n+1)$ Yang-Mills fields are retained can arise if we
consider a slightly enlarged higher-dimensional theory, now with a
dilatonic scalar as well as gravity and the $p$-form field strength.
Again, the key point is that an enhancement of the $GL(n,\R)$ global
symmetry that would occur for the reduction of a generic theory on
$T^n$ is needed, in order that the scalar coset manifold in the lower
dimension should have a denominator group that is large enough to
contain the desired $SO(n+1)$ local symmetry group of the theory
reduced on $S^n$.  It turns out that by including the dilatonic scalar
in the higher dimension, the necessary symmetry enhancements can be
achieved when the $p$-form field strength is either a 2-form, or a
3-form.  This opens up the possibility of finding consistent
Kaluza-Klein reductions on $S^2$, for the 2-form case, and on $S^3$ or
on $S^{D-3}$, for the 3-form case.  These, together with the previous
$D=11$ and $D=10$ examples, would then constitute the complete list 
of possibilities for consistent Kaluza-Klein $n$-sphere reductions
within the class of theories we are considering, in which all the
Yang-Mills fields of $SO(n+1)$ are retained.

    In this paper, we construct the complete and explicit non-linear
Kaluza-Klein Ans\"atze for these three new possibilities.  We begin in
section 2 with a detailed discussion of the global symmetry
enhancements that can occur in the toroidal reductions of theories
with gravity, a $p$-form field strength and a dilaton, in order to
establish what are the possibilities for consistent sphere reduction.
In section 3 we construct the Ansatz for the consistent reduction of
gravity plus a 3-form and a dilaton on $S^3$, keeping all the gauge
fields of $SO(4)$ and the ten scalars of the symmetric tensor
$T_{ij}$, together with the 2-form potential $A_\2$.  In section 4, we
examine two truncations of the $S^3$ reduction, in which only certain
subsets of the massless fields are retained, in order to make contact
with previous results in the literature.  In section 5 we consider the
``dual'' of the $S^3$ reduction namely, the reduction instead on
$S^{D-3}$.  Again, we find a consistent reduction Ansatz, in which all
the gauge fields of $SO(D-2)$ are retained, together with the
$\ft12(D-1)(D-2)$ scalars in $T_{ij}$.  A case of particular interest
is the $S^7$ reduction from $D=10$, since then the resulting
three-dimensional theory is the bosonic sector of a gauged $SO(8)$
supergravity, of a type not previously constructed in the literature.
In section 6 we construct the consistent Kaluza-Klein Ansatz for the
reduction of a theory of gravity, a dilaton and a 2-form field
strength on $S^2$.  In this case the Ansatz includes all three gauge
fields of $SO(3)$, together with the six scalars in $T_{ij}$.
  
    Note that in two of the new cases that we consider here, namely
the $S^2$ reduction of the theory with a 2-form field strength, and
the $S^{D-3}$ reduction of the theory with a 3-form field strength,
the totality of massless fields in the Kaluza-Klein reduced
lower-dimensional theories comprise the metric, the Yang-Mills gauge
fields $A_\1^{ij}$, and the scalars $T_{ij}$.  Thus these new examples
of consistent sphere reductions are akin to the $S^5$ reduction of
gravity plus a self-dual 5-form, in that no additional massless fields
are present that must also be included in the reduction Ansatz.  By
constrast, in the new $S^3$ reduction that we construct here, we must
additionally include the 2-form potential $A_\2$ in the Ansatz.  This
is similar to the situation for the $S^4$ and $S^7$ reductions, where,
as we discussed previously, additional massless fields are present,
and must be included, for consistency.

\section{Possibilities for $SO(n+1)$ Kaluza-Klein reductions on $S^n$}

    As we mentioned in the Introduction, a consistent Kaluza-Klein
reduction on $S^n$ that retains all the gauge fields of $SO(n+1)$ will
be possible only if there is a suitable enhancement of the generic
$GL(n,\R)$ global symmetry that arises instead in a $T^n$ reduction,
so that the denominator group of the generic scalar manifold
$GL(n,\R)/O(n)$ becomes large enough to contain $SO(n+1)$.  We also
mentioned that there are only rather limited circumstances under which
these symmetry enhancements can occur.

    The reason why the possibilities for symmetry enhancements are so
restrictive is discussed extensively in \cite{cjlp1,cjlp3}.  The
scalars divide into $n$ ``dilatons,'' $\vec\phi$ coming from the
diagonal metric components of the internal $n$-torus, with the rest
being ``axions'' $\chi_i$ coming from the off-diagonal metric
components and the reduction of the antisymmetric tensor.  Each
dilaton has a kinetic term of the form $-\ft12 e^{\vec c_i\cdot
\vec\phi}\, (\del\chi_i)^2$, where $\vec c_i$ is the associated
constant ``dilaton vector'' characterising the coupling of the
dilatons to that particular axion.

    In the $n$-torus reduction of a theory of $D$-dimensional gravity
plus $p$-form field strength with general values of $D$ and $p$, the
global symmetry will be $GL(n,\R)$.  In fact the dilaton vectors $\vec
b_i$ associated with the axions coming from the metric form the
complete set of positive roots of the $SL(n,\R)$ algebra
\cite{cjlp1,cjlp3}.  The dilaton vectors $\vec a_i$ associated with
the axions coming from the $p$-form field strength then form the
weights of some representation under $SL(n,\R)$.  If an enhancement of
the global symmetry is to occur, it must be that some or all of the
dilaton vectors $\vec a_i$ ``conspire'' to become the additional
positive roots of the enhanced symmetry algebra.  However, this cannot
occur in general, because the lengths of vectors $\vec a_i$ coming
from the $p$-form will be incommensurate with the lengths of the
vectors $\vec b_i$ coming from the metric.\footnote{We first, of
course, establish a canonical normalisation for the dilaton kinetic
terms, so that comparisons of the lengths are meaningful.}

     A convenient way to characterise the lengths of the various
dilaton vectors was introduced in \cite{stainless}.  Rather than using
the quantity $|\vec c\,|^2$ itself, it is convenient to introduce a
constant $\delta$, related to $|\vec c\,|^2$ by
\be
|\vec c\,|^2 = \delta - \fft{2(m-1)(D-m-1)}{D-2}\,,\label{cdelta}
\ee
where $D$ is the spacetime dimension, and $m$ is the degree of the
field-strength whose dilaton coupling is $e^{\vec c\cdot\vec\phi}$.
(Note that the all the field strengths in the $D=11$ and $D=10$
supergravities have $\delta=4$ couplings \cite{dkl}.)  The key point
about this parameterisation is that $\delta$ is preserved under
toroidal Kaluza-Klein reduction.  This makes it rather easy to see
when the possibility of an enhancement of the global symmetry can
occur.  First, we note \cite{stainless} that the dilaton vectors $\vec
b_i$ associated with the axions coming from the Kaluza-Klein reduction
of the metric {\it always} have $\delta=4$.  It follows therefore that
if the dilaton vectors $\vec a_i$ associated with axions coming from
the $p$-form field strength in the higher dimension are to have the
same lengths as the $\vec b_i$, then the coupling of the field
strength must also have $\delta=4$.\footnote{Since all the $\vec a_i$
themselves have equal length, and all the $\vec b_i$ themselves have
equal length, it follows that to get a {\it simply-laced} enhanced
symmetry group we would need that the length of the $\vec a_i$ and the
length of the $\vec b_i$ should be equal.  This turns out to be the
only situation where relevant symmetry enhancements occur, within the
framework of the higher-dimensional Lagrangians (\ref{dilatonlag}).
Thus although the case $D=6$ with a $\delta=2$ self-dual 3-form gives
an enhancement to $O(3,4)$ after a $T^3$ reduction to $D=3$
\cite{cjlp3}, and $D=5$ with a $\delta=\ft43$ 2-form gives an
enhancement to $G_2$ after a $T^2$ reduction to $D=3$ \cite{mo,cjlp3},
neither of these non-simply-laced cases would seem to indicate the
possibility of consistent $S^3$ or $S^2$ reductions.}  If we are
starting in $D$ dimensions with a theory with just gravity and the
$p$-form field strength, but no dilaton, this means that in $D$
dimensions we have $\vec c=0$, and $m=p$, and so to have $\delta=4$ we
must have
\be
2D-4 = (p-1)(D-p-1)\,.
\ee
It is easily verified that the only integer solutions are
$(D,p)=(11,4)$, $(11,7)$ and $(10,5)$. 

   The possibilities for achieving the necessary enhancement of the
global $GL(n,\R)$ symmetry can be broadened considerably if a dilatonic
scalar is included in the original theory in $D$ dimensions, since now
there is the possibility of adjusting its coupling to the $p$-form
field strength so that the corresponding value of $\delta$ is equal to
4.  Thus we may consider the $D$-dimensional Lagrangian
\be
{\cal L}_D = \hat R\, {\hat *\oneone} - \ft12 {\hat *d\hat\phi}\wedge
d\hat\phi - \ft12 e^{-a\, \hat\phi}\, {\hat *\hat F_{\sst{(p)}}}\wedge
\hat F_{\sst{(p)}}\,,\label{dilatonlag}
\ee
with $a$ chosen so that
\be
a^2 = 4 -\fft{2(p-1)(D-p-1)}{D-2}\,.\label{aval0}
\ee
Note that we put hats on all the fields in (\ref{dilatonlag}), to
indicate that they are higher-dimensional quantities.

    The first point to notice is that the requirement that the constant
$a$ should be real\footnote{One might in principle consider also the
possibility that $a$ could be imaginary.  This would be equivalent to
having a ghost-like kinetic term for the dilaton in the
$D$-dimensional theory.  This could not lead to any useful global symmetry
enhancements from the point of view of sphere reductions that retain
the $SO(n+1)$ gauge fields.  There might be possible implications for 
consistent reductions on spaces with non-compact symmetry groups.} 
is a rather restrictive one, since it implies 
\be
p^2-3D\, p + 3D - 5 \ge0\,.
\ee
Taken together with the fact that obviously $p$ cannot exceed $D$,
this implies that the only additional possibilities opened up by the
inclusion of the dilaton are for $p=2$, 3, $(D-2)$ or $(D-3)$.  The
last two here are just the Hodge duals of $p=2$ and $p=3$, so we need
not consider them as distinct cases.  For $p=2$ and $p=3$,
the relation (\ref{aval0}) gives
\bea
p=2:&&\qquad a^2 = \fft{2(D-1)}{D-2}\,,\label{p2case}\\
&&\nn\\
p=3:&&\qquad a^2 = \fft{8}{D-2}\,.\label{p3case}
\eea

    Thus we see that if we start in $D$ dimensions with the Lagrangian
(\ref{dilatonlag}) with a 2-form or 3-form field strength, we can
achieve a $\delta=4$ dilaton coupling in any dimension $D$, and thus
we can expect to find an enhancement of the $GL(n,\R)$ global symmetry
after dimensional reduction on $T^n$.  Indeed this is the case.  

    First, let us consider the case $p=3$, where we make a $T^n$
reduction of (\ref{dilatonlag}) with $a$ given by (\ref{p3case}).  The
global symmetry is indeed enhanced, and the scalar manifold in
$(D-n)$ dimensions will be \cite{sen1,sen2,cjlp3}
\bea
D-n>3:&&\qquad \R\times \fft{O(n,n)}{O(n)\times O(n)}\,,
\label{p3man}\\
&&\cr
D-n=3:&&\qquad \fft{O(D-2,D-2)}{O(D-2)\times O(D-2)}\,. \label{p3d3man} 
\eea 
This $p=3$ case corresponds precisely to
the T-duality symmetry of the toroidally-reduced bosonic string.
Note that if $D-n=3$ the usual T-duality group
$O(D-3,D-3)$ of the string theory reduced on $T^n$ is further enhanced
to the non-perturbative U-duality group $O(D-2,D-2)$.
Using the 3-form field strength, we can then consider either an $S^3$
or an $S^{D-3}$ Kaluza-Klein reduction.  

    If we take $n=3$, we see that the the scalar coset manifold from a
$T^3$ reduction will be
\be 
\R\times \fft{O(3,3)}{O(3)\times O(3)}\sim \R\times
\fft{SL(4,\R)}{SO(4)}\,.  
\ee 
There will also be six gauge potentials
coming from the Kaluza-Klein reduction on $T^3$.  This implies that
the $SO(4)$ subgroup of the $SL(4,\R)$ global symmetry group can be
gauged, with the six vector potentials becoming the Yang-Mills fields
of $SO(4)$. It is then natural to conjecture that this gauged theory
may be directly obtainable as a Kaluza-Klein reduction on $S^3$.  It
is far from obvious that such a reduction would be consistent,
since unlike the toroidal reduction there is no obvious
group-theoretic argument that would guarantee the consistency at the
non-linear level.\footnote{If we were gauging only the left-acting
$SU(2)$ or only the right-acting $SU(2)$ of the $SO(4)\sim
SU(2)_L\times SU(2)_R$ isometry of the 3-sphere (which is itself the
group manifold $SU(2)$), then the consistency would be guaranteed,
since the retained fields would then all be singlets under the {\it
other} $SU(2)$, but this is no longer the case when the gauge fields
of the full isometry group are retained.  In fact we shall discuss the
truncation to a single $SU(2)$ subgroup in section 4.}  In the next
section, we shall explicitly show that the reduction on $S^3$, in
which the full set of $SO(4)$ gauge fields are retained, is in fact
consistent at the full non-linear level.
 
   Now let us consider instead the $T^{D-3}$ reduction of (\ref{dilatonlag}),
again with $p=3$. The reduced theory will now be in three dimensions,
and the scalar coset manifold will be given by (\ref{p3d3man}),
provided that $a$ satisfies (\ref{p3case}).  Note that the further
symmetry enhancement of this $D-n=3$ case occurs because the complete
field content of the resulting three-dimensional theory (except for
the metric) can be described by scalars, since in three dimensions one
can dualise all the vector potentials to scalars.  The coset
(\ref{p3d3man}) can also be described as \cite{cjlp1}
\be
\fft{GL(D-2,\R)}{O(D-2)} \semi V\,,\label{rep1}
\ee
where $V$ is an irreducible representation under $GL(D-2,\R)$ of
dimension $\ft12(D-2)(D-3)$; this is the same as the dimension of the
adjoint representation of $O(D-2)$.

   The scalars in the representation $V$ can be dualised to vector
potentials,\footnote{The description (\ref{rep1}) would arise
naturally if one dualised the $(D-3)$ vector potentials coming from
the direct reduction of the original potential $\hat A_\2$, but left
all other vector potentials (including the Kaluza-Klein vectors) in
their original undualised forms.} suggesting that the $O(D-2)$
denominator group in (\ref{rep1}) can be gauged.  Then we may
conjecture that this gauged three-dimensional theory can alternatively
be obtained as a reduction of the original $D$-dimensional theory on
the sphere $S^{D-3}$.  In section 5, we shall demonstrate that there
is indeed such a consistent reduction on $S^{D-3}$, in which all the
gauge fields $A_\1^{ij}$ of $O(D-2)$ are retained, together with
scalars described by the symmetric tensor $T_{ij}$, where $i$ is a
vector index of $O(D-2)$.

    Finally, let us consider the Lagrangian (\ref{dilatonlag}) with
$p=2$, where the dilaton coupling for the 2-form is given by
(\ref{p2case}).  The Lagrangian (\ref{dilatonlag}) is then in fact
precisely the $S^1$ dimensional reduction of pure gravity in $D+1$
dimensions.  Consequently, the scalar manifold in $(D-n)$ dimensions
after reducing (\ref{dilatonlag}) on $T^n$ will be enhanced to
\be
\fft{GL(n+1,\R)}{SO(n+1)}\,.\label{p2man}
\ee
With a 2-form field strength we have in principle two possibilities
for sphere reductions, namely on $S^2$ or on $S^{D-2}$.  The latter
would be somewhat degenerate, since the lower-dimensional theory would
be in $D=2$, so we shall just consider the $S^2$ possibility here.  If
we take $n=2$, the denominator group in (\ref{p2man}) is exactly what
is needed to allow an $SO(3)$ gauging.  We may then conjecture that
this gauged theory should alternatively be directly obtainable as a
consistent Kaluza-Klein reduction on $S^2$, keeping all three of the
$SO(3)$ Yang-Mills gauge fields, together with six scalars described
by the symmetric tensor $T_{ij}$.  We shall in fact construct this
consistent reduction in section 6.

   We conclude this section with a Table that summarises all the cases
where consistent sphere reductions of a $D$-dimensional theory
comprising gravity, a $p$-form field strength and (in some cases) a
dilatonic scalar, are possible.  In all cases, we are concerned with
the situation where all the Yang-Mills fields of the $SO(n+1)$
isometry group of the $n$-sphere can be included in the reduction
Ansatz.

\bigskip\bigskip
\centerline{
\begin{tabular}{|c|c|c|c|c|c|c|}\hline
$p$-form & Dilaton & Higher-Dim & Lower-Dim. & Sphere & Gauge
Group &Extra fields\\ \hline\hline 
$F_\2$ & Yes & Any $D$ & $D-2$ & $S^2$ & $SO(3)$ &None \\ \hline 
$F_\3$ & Yes & Any $D$ & $D-3$ & $S^3$ & $SO(4)$ & $A_\2$\\ \hline
$F_\3$ & Yes & Any $D$ & $3$ & $S^{D-3}$ & $SO(D-2)$&None \\ \hline 
$F_\4$ & No & 11 & 7 & $S^4$ & $SO(5)$ &$A_\3^i$\\ \hline 
$F_\4$ & No & 11 & 4 & $S^7$ & $SO(8)$ &$\phi_{[ijk\ell]_+}$\\ \hline 
$F_\5={*F_\5}$ & No & 10 & 5 & $S^5$ & $SO(6)$ &None\\ \hline
\end{tabular}}
\bigskip

\noindent{\bf Table 1:} 
The possible cases for Kaluza-Klein $S^n$ reduction with
$SO(n+1)$ gauge fields.  The last column indicates what additional
fields, beyond the metric, the gauge fields $A_\1^{ij}$ and the
scalars $T_{ij}$, are massless, and must therefore be included, in a
consistent truncation (see discussion in section 1). 

\bigskip\bigskip

\section{Consistent $S^3$ reduction}

    We start from the bosonic string in $D$ dimensions, with the
low-energy effective Lagrangian\footnote{Later, in section 7, we shall
include the cosmological type term $-\ft12 m^2\, (D-26) \,
e^{\fft12a\, \hat\phi}$ that arises when $D\ne 26$, as a result of the
conformal anomaly.  For now, we restrict attention to the purely
classical Lagrangian for gravity coupled to a 3-form and a dilaton.}  
\be
{\cal L}_D = \hat R\, {\hat *\oneone} - 
\ft12 {\hat *d\hat\phi}\wedge d\hat\phi  
-\ft12 e^{-a\, \hat\phi}\, {\hat *\hat F_\3}\wedge \hat F_\3\,,
\label{dlag}
\ee
where the positive constant $a$ is given by (\ref{p3case}) so that the
global symmetry from a $T^n$ reduction would be $\R\times O(n,n)$
rather than merely $GL(n,\R)$, as discussed in section 2.  As we
argued there, we can now conjecture that it should be possible to
perform a consistent Kaluza-Klein reduction on $S^3$, keeping all the
$SO(4)$ Yang-Mills fields $A_\1^{ij}$, together with the scalar fields
described by the symmetric tensor $T_{ij}$, where $i$ is a vector
under $SO(4)$, and also the 2-form potential $A_\2$.

    We find that there is indeed an Ansatz for a consistent 
Kaluza-Klein reduction on $S^3$, given by
\bea
&&d\hat s_D^2 = Y^{\fft1{D-2}}\, \Big( \Delta^{\fft2{D-2}}\, ds_{D-3}^2 
+ g^{-2}\, \Delta^{-\fft{D-4}{D-2}}\,  T^{-1}_{ij}\, 
\cD\mu^i\, \cD\mu^j\Big) \,,\label{metans}\nn\\
&&e^{\sqrt{(D-2)/2}\, \hat\phi} = \Delta^{-1}\, Y^{(D-4)/4}\,,
\label{phians}\nn\\
&&\hat F_\3 = F_\3 + \ft16\, \ep_{i_1 i_2 i_3 i_4}\, 
\Big( g^{-2}\, U\, \Delta^{-2}\, 
  \cD\mu^{i_1}\wedge \cD\mu^{i_2} \wedge \cD\mu^{i_3}\, 
\mu^{i_4} \label{fans}\\
&&\quad- 3g^{-2}\, \Delta^{-2} \, 
D\mu^{i_1} \wedge \cD\mu^{i_2}\wedge \cD T_{i_3 j}\,
T_{i_4 k}\, \mu^j\, \mu^k  - 3g^{-1}\, \Delta^{-1}\, F_\2^{i_1 i_2} \wedge
\cD\mu^{i_3}\, T_{i_4 j}\, \mu^j \Big)\,,\nn
\eea
where
\bea
&&\mu^i \, \mu^i = 1\,,\qquad \Delta = T_{ij}\, \mu^i\, \mu^j\,,\qquad
U = 2 T_{ik}\, T_{jk}\, \mu^i\, \mu^j - \Delta \, T_{ii}\,,\nn\\
&& Y = \det(T_{ij})\,,\label{somedefs}
\eea
and the indices $i,j,\ldots$ range of 4 values.  Here, and in the rest
of the paper, a summation over repeated $SO(n+1)$ indices is understood.
The gauge-covariant exterior derivative $D$ is defined so that
\be
\cD\mu^i = d\mu^i + g\, A_\1^{ij}\, \mu^j\,,\qquad
\cD T_{ij} = dT_{ij} + g\, A_\1^{ik}\, T_{kj} + g\, A_\1^{jk}\, T_{ik}\,,
\label{gaugecov}
\ee
where $A_\1^{ij}$ denotes the $SO(4)$ gauge potentials coming from the
isometry group of the 3-sphere, and
\be
F_\2^{ij} = dA_\1^{ij} + g\, A_\1^{ik}\wedge A_\1^{kj}\,.
\label{fieldstrength}
\ee
Thus the lower-dimensional fields appearing in the Kaluza-Klein Ansatz
comprise the metric $ds_{D-3}^2$, the six gauge potentials $A_\1^{ij}$
of $SO(4)$, the ten scalar fields described by the symmetric tensor
$T_{ij}$, and the 2-form potential $A_\2$, whose (Chern-Simons
modified) field strength is $F_\3$.  The form of the Ansatz that we
have obtained here closely parallels the structure found in
\cite{clpst} for the $S^5$ reduction of type IIB supergravity.

   In order to demonstrate the consistency of the Kaluza-Klein
reduction with the above Ansatz, we substitute it into the
$D$-dimensional equations of motion\footnote{We shall not consider the
reduction of the $D$-dimensional Einstein equation in detail in this
paper, on account of its complexity; this will be addressed in future
work. In practice, in all cases that have been examined, the Einstein
equation seems always to give consistent results provided that the
equations of motion for all the other fields are consistent.
Furthermore, the agreement of our reduction Anstaz with
previously-established special cases provides further supporting
evidence for the consistency of the Einstein equation.}  that follow
from (\ref{dlag}).  These are
\bea
d{\hat *d\hat\phi} &=& -\ft12 a\, e^{-a\, \phi}\, \hat F_\3\wedge {\hat
*\hat F_\3}\,,\nn\\
d(e^{-a\, \phi}\, {\hat * \hat F_\3}) &=& 0\,,\label{deom}\\
\hat R_{MN} &=& \ft12 \del_M\phi\, \del_N\phi + \ft14\, \Big( \hat
F^2_{MN} - \fft{2}{3(D-2)}\, \hat F_\3^2\, \hat g_{MN}\Big)\,.\nn
\eea
In addition, we have the Bianchi identity $d\hat F_\3=0$.  Taking this
first, we find upon substituting $\hat F_\3$ from (\ref{fans}) into
$d\hat F_\3=0$ that the lower-dimensional field $F_\3$ must satisfy
the Bianchi identity 
\be
dF_\3 = \ft18 \ep_{i_1 i_2 i_3 i_4}\, F_\2^{i_1 i_2}\wedge F_\2^{i_3
i_4}\,.\label{f3bianchi}
\ee
All other terms arising from $d\hat F_\3$ vanish identically.
(The calculation is quite involved, and the Schoutens identity
$\ep_{[i_1 i_2 i_3 i_4}\, V_{i_5]}\equiv 0$ is useful.) 

   In order to substitute the Ansatz into the field equation for $\hat
F_\3$, we must first calculate the Hodge dual of $\hat F_\3$ given in
(\ref{fans}).  This is a straightforward, although somewhat involved
calculation, and we find
\bea
e^{-\sqrt{8/(D-2)}\, \hat\phi}\, {\hat * \hat F_\3} &=& \fft1{6g^3}\,
\ep_{ijk\ell}\,  Y^{-1} {* F_\3}\wedge \mu^i\, \cD\mu^j\wedge \cD\mu^k
\wedge \cD \mu^\ell -g\, U\, \ep_{D-3}\label{fdual}\\
&& + g^{-1}\, T^{-1}_{ij}\, 
{*\cD T_{jk}}\wedge (\mu^k\, \cD\mu^i) 
-\fft1{2g^2}\,  T^{-1}_{ik}\, T^{-1}_{jl}\, 
{* F_\2^{ij}}\wedge \cD\mu^k\wedge
\cD\mu^\ell\,.\nn
\eea
Substituting into (\ref{deom}), with $a$ given by (\ref{p3case}), we
(eventually) read off the lower-dimensional equations of motion
\bea
&&(-1)^D\, \cD(T^{-1}_{ik}\, T^{-1}_{j\ell}\, {*F_\2^{k\ell}}) = 
-2g\, T^{-1}_{k[i}\, {*\cD}T_{j] k} -\ft12 \ep_{ijk\ell}\,  
Y^{-1}\, {*F_\3}\wedge 
F_\2^{k\ell}\,,\nn\\
&&(-1)^D\, \cD(\wtd T^{-1}_{ik}\, {*\cD} \wtd T_{kj}) =  
2g^2\, [2T_{ik}\, T_{jk} - T_{ij} \,
T_{kk}] \, \ep_{D-3} - T^{-1}_{\ell m}\, T^{-1}_{ik}\,
{*F_\2^{\ell k}}\wedge F_\2^{mj}\nn\\
&& \phantom{xxxxxxxxxxxxxx}
- \ft14\delta_{ij}\, \Big(2g^2\, [2T_{n k}\, T_{n k} -
(T_{kk})^2] \, \ep_{D-3} -T^{-1}_{\ell m}\, T^{-1}_{nk}\,
{*F_\2^{\ell k}}\wedge F_\2^{mn}\Big)\,,\nn\\
&&d(Y^{-1}\, {*F_\3}) =0\,.
\eea
We have introduced the unimodular matrix $\wtd T_{ij}$,
constructed from $T_{ij}$ by extracting the determinant factor $Y$
(see (\ref{somedefs})),
\be
\wtd T_{ij} = Y^{\fft14}\, T_{ij}\,.
\ee
Again, there are many other terms that arise
from acting with the exterior derivative that cancel amongst
themselves, after making use of the Schoutens identity.  The
consistency of the reduction Ansatz manifests itself in the remarkable
fact that one reads off consistent lower-dimensional equations of
motion in which all the dependence on the internal $S^3$ coordinates
$\mu^i$ has cancelled.

   Next, we consider the equation of motion for the dilaton $\hat\phi$.  
 From (\ref{phians}) we find
\be
d\hat\phi = \sqrt{\fft2{D-2}}\, \Big(\ft14(D-4)\, Y^{-1}\, {*dY} -
\Delta^{-1}\, d\Delta\Big)\,.
\ee
Since $\Delta$ has dependence on the $S^3$ coordinates $\mu^i$ one of
the terms arising here will involve the quantity
\be
T_{ij}\, \mu^i\, \cD\mu^j\,.
\ee
It is therefore necessary to evaluate the Hodge dual of this 1-form; we find
\be
{\hat *(T_{ij}\, \mu^i\, \cD\mu^j)} = -\ft12 \ep_{i_1 i_2 i_3 i_4}\, 
T_{i\ell}\, \mu^i\, \ep_{D-3}\wedge\, ( \Delta\, T_{i_1 \ell} -
T_{i_1 j}\, T_{k \ell}\, \mu^j\, \mu^k) \, \mu^{i_2}\,
\cD\mu^{i_3}\wedge \cD\mu^{i_4}\,.
\ee
After some involved manipulations, we find that the $D$-dimensional
dilaton equation of motion in (\ref{deom}) implies that $Y$ satisfies
\bea
\ft{D-5}{4}\, (-1)^D\, d(Y^{-1}\, {*dY}) &=& \ft12 g^2\, 
(2T_{ij}\, T_{ij} -(T_{ii})^2)\,
\ep_{D-3} - Y^{-1}\, {* F_\3}\wedge F_\3 \nn\\
&&- \ft14 Y^{-1}\,
T^{-1}_{ik}\, T^{-1}_{j\ell}\, {*F_\2^{ij}}\wedge F_\2^{k\ell}\,.
\eea

     The full set of $(D-3)$-dimensional equations of motion can be
derived from the Lagrangian
\bea
{\cal L}_{D-3} &=& R\, {*\oneone} - \ft{D-5}{16}\, Y^{-2}\,
{*dY}\wedge dY - \ft14 \wtd T^{-1}_{ij}\, {*\cD \wtd T_{jk}}\wedge
\wtd T^{-1}_{k\ell}\, \cD\wtd T_{\ell i} \nn\\
&& - \ft12 Y^{-1}\, {*F_\3}\wedge
F_\3 -\ft1{4}\, Y^{-\fft12}\,  \wtd T^{-1}_{ik}\,
\wtd T^{-1}_{j\ell}\, {* F_\2^{ij}}\wedge F_\2^{k\ell}
-V\, {*\oneone}\,,\label{dm3lag}
\eea
where the potential $V$ is given by
\be
V = \ft12 g^2\, Y^{\fft12}\, \Big(2 \wtd T_{ij}\, \wtd T_{ij} 
  - (\wtd T_{ii})^2 \Big)\,.\label{s3pot}
\ee
The 3-form field strength $F_3$ is given by
\be
F_\3 = dA_\2 +\ft18 \ep_{ijk\ell}\, (F_\2^{ij}\wedge A_\1^{k\ell}
-\ft13 g\, A_\1^{ij}\wedge A_\1^{km}\wedge A_\1^{m\ell})\,,
\ee
which implies that $F_\3$ satisfies the Bianchi identity
(\ref{f3bianchi}). 

\section{Truncations to previous results}

    In this section, we consider two truncations of the $S^3$
Kaluza-Klein reduction of the bosonic string that we constructed in
the previous section.  

\subsection{Truncation from $SO(4)$ to $SU(2)$} 

The first truncation turns the reduction into a
``standard'' one, for which the consistency becomes immediately
understandable from group-theoretic arguments.  Specifically, we may
truncate the $SO(4)$ Yang-Mills gauge fields that arise from the $S^3$
reduction to a set of $SU(2)$ gauge fields, corresponding either to
the left-action, or to the right-action, of $SU(2)$ on the $S^3\sim
SU(2)$ group manifold.  This is achieved by imposing a self-dual or
anti-self-dual truncation on the original $SO(4)$ gauge potentials
$A_\1^{ij}$, 
\be
A_\1^{ij} = \pm \ft12 \ep_{ijk\ell}\, A_\1^{k\ell}\,.\label{selfdual}
\ee
The choice of sign governs whether we are retaining the gauge fields
of $SU(2)_L$ or of $SU(2)_R$ in the truncation of $SO(4)\sim
SU(2)_L\times SU(2)_R$.  The two choices are equivalent, up to
convention choices, and we shall pick the plus sign in
(\ref{selfdual}) for definiteness.  It is convenient to take the $i,j,\ldots$
indices to range over the values $0,1,2,3$, and to write the remaining
gauge potentials in terms of the $SU(2)$ triplet $A_\1^\a$, with
\be
A_\1^{01}=A_\1^{23}\equiv \ft12 A_\1^1\,,\qquad
A_\1^{02}=A_\1^{31}\equiv -\ft12 A_\1^2\,,\qquad
A_\1^{03}=A_\1^{12}\equiv \ft12 A_\1^3\,.
\ee
These are the gauge fields of $SU(2)_L$.

    At the same time as we impose the self-dual truncation
(\ref{selfdual}) on the gauge potentials, we must also truncate the
scalar fields $T_{ij}$, in order to be consistent with the equations
of motion for the truncated gauge fields.  In fact we should retain
just a single scalar degree of freedom $X$, so that $T_{ij}$ becomes
\be
T_{ij} = X\, \delta_{ij}\,.
\ee
Note that from (\ref{somedefs}) we shall now have $Y=X^4$.  It is
convenient also to give an explicit parametrisation of the $\mu^i$
coordinates, in terms of Euler angles on $S^3$:
\be
\mu_0+ \im\,  \mu_3 = \cos\ft12\theta\, e^{\im\,
(\psi+\phi)/2}\,,\qquad
\mu_1+ \im\,  \mu_2 = \sin\ft12\theta\, e^{\im\,
(\psi-\phi)/2}\,.
\ee
In terms of these we can then define the left-invariant 1-forms
$\sigma_\a$ on $S^3$, according to
\be
\sigma_1+\im\, \sigma_2 = e^{-\im\, \psi}\, (d\theta + \im\,
\sin\theta\, d\phi)\,,\qquad \sigma_3= d\psi + \cos\theta\, d\phi\,.
\ee
These satisfy the $SU(2)$ algebra $d\sigma_\a = -\ft12
\ep_{\a\beta\gamma}\, \sigma_\beta \wedge \sigma_\gamma$.  

    With these preliminaries, we can now present our results for the
reduction Anstaz for this $SU(2)$ truncation of the original
$SO(4)$ Kaluza-Klein reduction.   We find that the metric, dilaton and
3-form Ans\"atze given in (\ref{metans})-(\ref{fans}) reduce to
\bea
d\hat s_D^2 &=& X^{\fft{6}{D-2}}\, ds_{D-3}^2 + 
\ft14 X^{-\fft{2(D-5)}{D-2}}\, \sum_\a 
(\sigma_\a-g\, A_\1^\a)^2\,,\label{su2metans}\\
e^{\sqrt{(D-2)/2}\, \hat\phi} &=& X^{D-5}\,,\label{su2phians}\\
\hat F_\3 &=& F_\3 - \fft1{4g^2}\, \Omega_\3 - \fft1{12 g}\,
\ep_{\a\beta\gamma}\, F_\2^\a\wedge (\sigma_\beta -g \,
A_\1^\beta)\wedge (\sigma_\gamma - g\, A_\1^\gamma)\,,\label{su2fans}
\eea
where 
\be
\Omega_\3\equiv \ft16 \ep_{\a\beta\gamma}\, 
(\sigma_\a-g\, A_\1^\a)\wedge (\sigma_\beta-g\, A_\1^\beta)\wedge
(\sigma_\gamma-g\, A_\1^\gamma)
\ee
is the volume form on the 3-sphere.

   It is easy to verify that this $SU(2)$ truncation of the full
$SO(4)$ reduction Ansatz of section 2 is a consistent one.  As we
remarked above, there is no longer anything ``surprising'' about the
consistency in this case, since the truncation has set to zero all
fields that transformed non-trivially under $SU(2)_R$.  In other
words, the $SU(2)_L$ Ansatz in this section retains all the singlets
under $SU(2)_R$, while discarding all the non-singlets.  Such a
truncation is necessarily consistent, since non-linear products of the
fields that are retained can never generate non-singlets under
$SU(2)_R$.  A related point is that the fields that remain in the
reduction Ansatz parameterise {\it homogeneous} deformations of the 3-sphere.
A particular case of this $SU(2)$ reduction has appeared
previously in the literature, in the $S^3$ reduction of $N=1$
supergravity from $D=10$ to $D=7$ \cite{chamsab}.  

\subsection{Truncation from $SO(4)$ to $U(1)\times U(1)$}

    The second truncation that we shall consider here involves
retaining only the $U(1)\times U(1)$ subgroup of the original
$SU(2)\times SU(2)$ gauged fields of the full $SO(4)$ reduction Ansatz
of section 2. It is convenient now to take the $SO(4)$ indices
$i,j,\ldots$ to range over the values $1,2,3,4$.  The truncation
amounts to setting all gauge potentials $A_\1^{ij}$ to zero except for 
$A_\1^{12}$ and $A_\1^{34}$, for which we write
\be
A_\1^{12} \equiv A_\1^1\,,\qquad A_\1^{34} \equiv A_\1^2\,.
\ee
It is also convenient now to parameterise the coordinates $\mu^i$ on
$S^3$ as
\be
\mu_1+\im\, \mu_2 = \td\mu_1\, e^{\im\, \phi_1}\,,\qquad
\mu_3+\im\, \mu_4 = \td\mu_2\, e^{\im\, \phi_2}\,.
\ee
At the same time as making the truncation of the gauge fields,
consistency with their equations of motion requires that we set
certain of the scalar fields to zero, so that what remains is just two
scalars $X_1$ and $X_2$ as follows:
\be
T_{ij} = {\rm diag}\, (X_1, X_1, X_2, X_2)\,.
\ee
Note that we shall now have
\be
Y=(X_1\, X_2)^2\,,\qquad \Delta= X_1\, \td\mu_1^2 + X_2\, \td
\mu_2^2\,,\qquad
U = 2 \sum_{i=1}^2 (X_i^2\, \td\mu_i^2 - \Delta\, X_i)\,.
\ee

    After substituting the above truncation and reparametrisation into
the original Kaluza-Klein Ans\"atze in section 2, we find that the
metric and dilaton As\"atze become
\bea
d\hat s_D^2 &=& (X_1\, X_2)^{\fft2{D-2}}\, \Big(
\Delta^{\fft2{D-2}}\, ds_{D-3}^2\nn\\
&&\qquad + g^{-2}\, \Delta^{-\fft{D-4}{D-2}}\, 
\sum_{i=1}^2 X_i^{-1}\, (d\td\mu_i^2 + \td\mu_i^2\, (d\phi_i - g\,
A_\1^i)^2)\Big) \,,\label{u1u1metans}\\
e^{\sqrt{(D-2)/2}\, \hat\phi} &=& \Delta^{-1}\, (X_1\,
X_2)^{\fft{D-4}{2}}\,.\label{u1u1phians}
\eea
The Ansatz for the 3-form field $\hat F_\3$ in this $U(1)^2$
truncation is most simply expressed in terms of the expression for the
dual of $\hat F_\3$.  Making the truncation in (\ref{fdual}), we find
\bea
e^{-\sqrt{8/(D-2)}\, \hat\phi}\, {\hat * \hat F_\3} &=& 
-2g\, \sum_{i=1}^2 \Big( X_i^2\, \td\mu_i^2 - \Delta\, X_i\Big) 
 \ep_{D-3} +\fft1{2g}\, \sum_{i=1}^2 X_i^{-1}\, {*dX_i}\wedge 
d(\td\mu_i^2)\nn\\
&& \!\!\!\!\! - \fft1{2g^2}\, \sum_{i=1}^2 X_i^{-2}\, d(\td\mu_i^2)\wedge
(d\phi_i - g\, A_\1^i)\wedge {*F_\2^i}
+g^{-3}\, Y^{-1} {* F_\3}\,.\label{u1u1fdual}
\eea
The Ansatz for $\hat F_\3$ itself is also easily obtainable by
imposing the $U(1)^2$ truncation on the general $SO(4)$ Ansatz (\ref{fans}).

    Note that in the $U(1)^2$ truncation the question of the consistency of the
reduction is still a non-trivial one, since the two scalars $X_1$ and
$X_2$ parameterise inhomogeneous deformations of the 3-sphere.  Of
course since we have already argued that the $SO(4)$ reduction in
section 2 is consistent, the consistency for the $U(1)^2$ truncation
is a consequence.

    A particular case of this $U(1)^2$ truncation appeared previously
in the literature \cite{cllp}, where it was obtained for the case
$D=10$ by taking a singular limit of the $S^4$ reduction of
eleven-dimensional supergravity that was constructed in \cite{ten}.

\section{$S^{D-3}$ reduction and $D=3$, $N=8$ gauged supergravity}

     As we discussed in section 2, it is natural to conjecture that 
the theory of gravity coupled to a dilaton and a 3-form, described by
(\ref{dlag}) with $a$ given by (\ref{p3case}), should also admit a
consistent reduction to three dimensions on the sphere $S^{D-3}$, in
which all the Yang-Mills gauge fields of $SO(D-2)$ are retained.
Additionally, we should keep the $\ft12(D-1)(D-2)$ scalar fields
described by the symmetric tensor $T_{ij}$, where $i$ is a vector
index of $SO(D-2)$.  We find that indeed such a consistent reduction
is possible, and that the Kaluza-Klein Ansatz is given by
\bea
d\hat s_D^2 &=& Y^{\fft1{D-2}}\, \Big(\Delta^{\fft{D-4}{D-2}}\, ds_{3}^2 
+ g^{-2}\, \Delta^{-\fft2{D-2}}\,  T^{-1}_{ij}\, 
\cD\mu^i\, \cD\mu^j\Big)
\,,\label{metans3}\nn\\
e^{-\sqrt{(D-2)/2}\, \hat\phi} &=& \Delta^{-1}\, Y^{1/2}\,,
\label{phians3}\nn\\
 \hat F_\3 &=&
-g\, U\, \ep_{3} + g^{-1}\, T^{-1}_{ij}\, 
{*\cD T_{jk}}\wedge (\mu^k\, \cD\mu^i) \nn\\
&&
-\fft1{2g^2}\,  T^{-1}_{ik}\, T^{-1}_{jl}\, 
{* F_\2^{ij}}\wedge \cD\mu^k\wedge
\cD\mu^\ell\,,\label{fans3}
\eea
where the various quantities appearing here are again given in
(\ref{somedefs}), (\ref{gaugecov}) and (\ref{fieldstrength}), but now
the indices $i,j,\ldots$ range over $(D-2)$ values.  Thus the field
content in Kaluza-Klein reduced three-dimensional theory comprises the
metric $ds_3^2$, the $\ft12(D-2)(D-3)$ gauge potentials $A_\1^{ij}$ of
$SO(D-2)$, and the $\ft12 (D-1)(D-2)$ scalars described by the
symmetric tensor $T_{ij}$.  The calculation of the Hodge dual of the
3-form $\hat F_\3$ is again a mechanical, although involved,
calculation.  We find that it is given by
\bea
e^{-\sqrt{8/(D-2)}\, \hat\phi} \, {\hat * \hat F_\3} &=& 
\fft{g^{-(D-4)}}{(D-3)!}\, \ep_{i_1\cdots i_{D-2}}  \Big(
g\, U\, \Delta^{-2}\, \mu^{i_1}\, \cD\mu^{i_2}\cdots \cD\mu^{i_{D-2}}
\nn\\
&& - {\scriptstyle (D-3)}\, 
\Delta^{-2}\, T_{i_1 j}\, \cD T_{i_2 k}\, \cD\mu^{i_3}\cdots
\cD\mu^{i_{D-2}} \, \mu^j\, \mu^k\nn\\
&& -\ft{(D-3)(D-4)}{2}\, F_\2^{i_1 i_2}\, T_{i_3 j}\, \cD\mu^{i_4}\,
\cD\mu^{i_{D-2}}\, \mu^j \Big)\,,
\eea
where we have suppressed the wedge symbols in products of differential
forms in order to economise on space.

    It is again a straightforward, although lengthy, procedure to
substitute the above Ansatz into the $D$-dimensional equations of
motion (\ref{deom}), and to verify that there is a consistent
reduction to equations of motion for the three-dimensional fields. 
We find that these equations can be derived from the following
three-dimensional Lagrangian:
\bea
{\cal L}_3 &=& R\, {*\oneone} - \ft1{4(D-2)}\, Y^{-2}\,
{*dY}\wedge dY - \ft14 \wtd T^{-1}_{ij}\, {*\cD \wtd T_{jk}}\wedge
\wtd T^{-1}_{k\ell}\, \cD\wtd T_{\ell i} \nn\\
&& -\ft1{4}\, Y^{-\fft2{D-2}}\,  \wtd T^{-1}_{ik}\,
\wtd T^{-1}_{j\ell}\, {* F_\2^{ij}}\wedge F_\2^{k\ell}
-V\, {*\oneone}\,,\label{d3lag}
\eea
where $Y=\det(T_{ij})$, and $T_{ij}$ is written in terms of the
unimodular $(D-2)\times (D-2)$ matrix $\wtd T_{ij}$ as $T_{ij}=
Y^{1/(D-2)}\, \wtd T_{ij}$.  The potential $V$ is given by
\be
V = \ft12 g^2\, Y^{\fft2{D-2}}\, \Big(2 \wtd T_{ij}\, \wtd T_{ij} 
  - (\wtd T_{ii})^2 \Big)\,.\label{d3pot}
\ee

     An application of this dimensional reduction that is of
particular interest arises if we take $D=10$, since then the
higher-dimensional starting point will be the bosonic sector of $N=1$
supergravity in ten dimensions.  The reduction on $S^7$ then yields a
three-dimensional theory that is the bosonic sector of an $SO(8)$ gauged 
supergravity, with $N=8$ (\ie half of maximal) supersymmetry.  As well
as the 28 gauge fields, there are in total 36 scalars, described by
the unimodular symmetric tensor $\wtd T_{ij}$ and the scalar $Y$.
These transform as a 35 and a 1 under $SO(8)$, respectively.
Evidently, if we reduced the full $N=1$ theory in $D=10$, including
the fermions, we would obtain $N=8$ gauged $SO(8)$ supergravity in
three dimensions. This appears to be the first example of such a
gauged supergravity in $D=3$.  Previous examples of gauged
three-dimensional supergravities in the literature
have been of the type constructed in \cite{achutown}, with
$SO(p)\times SO(q)$ gauge fields and a pure cosmological constant term
implying the existence of an AdS$_3$ ground-state solution.  In fact
there are no scalar fields, and hence no scalar potential, in the theories
constructed in \cite{achutown}.  By contrast, the gauged supergravity
that we have obtained here has 36 scalars with the potential
(\ref{d3pot}).  The theory does not admit an AdS$_3$ solution, but it
may allow domain-wall solutions that preserve half of the supersymmetry.

\section{$S^2$ reduction}

   Here, we construct the Kaluza-Klein Ansatz for the reduction of
(\ref{dilatonlag}) with $p=2$ and $a$ given by (\ref{p2case}).  Thus
our starting point is
\be
{\cal L}_D = \hat R\, {\hat *\oneone} - 
\ft12 {\hat *d\hat\phi}\wedge d\hat\phi  
-\ft12 e^{-a\, \hat\phi}\, {\hat *\hat F_\2}\wedge \hat F_\2\,,
\label{dlagf2}
\ee
where the positive constant $a$ is given by (\ref{p2case}).
From (\ref{dlagf2}) we derive the equations of motion
\bea
d{\hat *d \hat\phi} &=& \ft12(-1)^D\, a\, e^{-a\, \hat\phi}\, {\hat
*\hat F_\2}\wedge \hat F_\2\,,\nn\\
d(e^{-a\, \hat\phi} {\hat *\hat F_\2}) &=& 0\,,\label{f2eom}\\
\hat R_{MN} &=& \ft12 \del_M\hat\phi\, \del_N\hat\phi + 
\ft12 e^{-a\,\hat\phi}\, \Big( \hat F^1_{MN} - \ft1{2(D-2)}\, \hat
F_\2^2\, \hat g_{MN}\Big)\,.
\eea

     We find that there is a consistent reduction Ansatz on $S^2$, given by
\bea
d\hat s_D^2 &=& Y^{\fft1{D-2}}\, \Big(\Delta^{\fft1{D-2}}\, ds_{D-2}^2 
+ g^{-2}\, \Delta^{-\fft{D-3}{D-2}}\, T^{-1}_{ij}\, 
\cD\mu^i\, \cD\mu^j\Big)
\,,\label{metanss2}\\
e^{\sqrt{\fft{2(D-2)}{D-1}}\, \hat\phi} &=& \Delta^{-1}\, 
Y^{\fft{D-3}{D-1}}\,, \label{phianss2}\\
\hat F_\2 &=& \ft12 \ep_{ijk}\, \Big( g^{-1}\, U\, \Delta^{-2}\, 
\mu^i\, \cD\mu^j\wedge \cD\mu^k - 2 g^{-1}\, \Delta^{-2}\,
\cD\mu^i\wedge \cD T_{j\ell}\, T_{k m}\, \mu^\ell\, \mu^m\nn\\
&& - \Delta^{-1}\, F_\2^{ij}\, T_{k\ell}\, \mu^\ell \Big)\,.
\label{f2ans}
\eea
Again, the various quantities appearing here are given in
(\ref{somedefs}), (\ref{gaugecov}) and (\ref{fieldstrength}), but with
the indices $i,j,\ldots$ ranging over 3 values.  
The dual of the 2-form then turns out to be given by
\bea
e^{-\sqrt{\fft{2(D-1)}{D-2}}\, \hat\phi}\, {\hat * \hat F_\2} &=& 
-g\, U\, \ep_{D-2} + g^{-1}\, T^{-1}_{ij}\, 
{*\cD T_{jk}}\wedge (\mu^k\, \cD\mu^i) 
\nn\\
&&-\fft1{2g^2}\,  T^{-1}_{ik}\, T^{-1}_{jl}\, 
{* F_\2^{ij}}\wedge \cD\mu^k\wedge
\cD\mu^\ell\,.\label{f2dual}
\eea
The field content of the Kaluza-Klein reduced theory comprises the
$(D-2)$-dimensional metric $ds_{D-2}^2$, the three gauge potentials
$A_\1^{ij}$ of $SO(3)$, and the six scalar fields $T_{ij}$.

    Substituting the Ansatz into the $D$-dimensional equations of
motion (\ref{f2eom}), we find that it yields a consistent Kaluza-Klein $S^2$
reduction, with the $(D-2)$-dimensional fields satisfying equations of
motion that follow from the Lagrangian
\bea
{\cal L}_{D-2} &=& R\, {*\oneone} - \ft{D-4}{3(D-1)}\, Y^{-2}\,
{*dY}\wedge dY - \ft14 \wtd T^{-1}_{ij}\, {*\cD \wtd T_{jk}}\wedge
\wtd T^{-1}_{k\ell}\, \cD\wtd T_{\ell i} \nn\\
&& -\ft1{4}\, Y^{-\fft23}\,  \wtd T^{-1}_{ik}\,
\wtd T^{-1}_{j\ell}\, {* F_\2^{ij}}\wedge F_\2^{k\ell}
-V\, {*\oneone}\,,\label{dm2lag}
\eea
where $Y=\det(T_{ij})$, and $T_{ij}$ is written in terms of the
unimodular $3\times 3$ matrix $\wtd T_{ij}$ as $T_{ij}= Y^{1/3}\, \wtd
T_{ij}$.  The potential $V$ is given by
\be
V = \ft12 g^2\, Y^{\fft23}\, \Big(2 \wtd T_{ij}\, \wtd T_{ij} 
  - (\wtd T_{ii})^2 \Big)\,.\label{dm2pot}
\ee

   In view of our earlier observation that the $D$-dimensional Lagrangian
(\ref{dlagf2}), with the constant $a$ given by (\ref{p2case}), can
itself be thought of as an ordinary $S^1$ Kaluza-Klein reduction of
pure gravity in $(D+1)$ dimensions, it follows that we can also
interpret our result as a consistent reduction of $(D+1)$-dimensional
pure gravity.  The internal space is not simply
$S^1\times S^2$, however, since the 2-form field $F_\2$ in $D$
dimensions, which is the Kaluza-Klein vector of the $S^1$ reduction
from $(D+1)$ dimensions, is topologically non-trivial.  One can see
from (\ref{f2ans}) that if, for example, the scalars were all taking
trivial values, the 2-form field $\hat F_\2$ would be just the
volume-form on $S^2$ (like in a Dirac monopole configuration).   Thus
the reduction from $(D+1)$ dimensions is actually on a manifold that
is topologically $S^3$.  In fact we can easily lift the metric Ansatz
in (\ref{metanss2}) to give the Ansatz for the reduction from $(D+1)$
dimensions, by incorporating the standard $S^1$ reduction step
\be
d\hat s_{D+1}^2 = e^{2\a\, \hat \phi}\, d\hat s_D^2 + e^{-2\a\,
(D-2)\, \hat \phi}\, (dz + \hat A_\1)^2\,,
\ee
where $\hat F_\2= d\hat A_\1$, and the fields on the right-hand side
are given in (\ref{metanss2})--(\ref{f2ans}).  Thus we find
\be
d\hat s_{D+1}^2 = Y^{\fft2{D-1}}\, ds_{D-2}^2 + \Delta^{-1}\,
Y^{\fft2{D-1}}\, T_{ij}^{-1}\, \cD\mu^i\, \cD\mu^j + \Delta\,
Y^{-\fft{D-3}{D-1}}\, (dz+ \hat A_\1)^2\,.
\ee
This is an unusual type of $S^3$ reduction, in which the three $SO(3)$
Yang-Mills fields $A_\1^{ij}$ and the six scalar fields $T_{ij}$
parameterise inhomogeneous deformations of the 3-sphere.

\section{Conformal anomaly terms}

    Until now we have focussed our attention on the purely classical
theories of gravity coupled to a $p$-form field strength and a
dilaton.  One of the two cases that admits consistent sphere
reductions turned out to be when $p=3$, and in fact the Lagrangian
(\ref{dlag}) is precisely the leading-order expression for the
low-energy limit of the $D$-dimensional bosonic string.  Of course the
bosonic string suffers from a conformal anomaly if the dimension $D$
is not equal to 26.  It turns out that the effect of this anomaly is
to generate an additional term in the effective action \cite{cmpf},
which vanishes at $D=26$, so that (\ref{dlag}) is replaced by
\be
{\cal L}_D = \hat R\, {\hat *\oneone} - 
\ft12 {\hat *d\hat\phi}\wedge d\hat\phi  
-\ft12 e^{-a\, \hat\phi}\, {\hat *\hat F_\3}\wedge \hat F_\3
-\ft12 m^2\,(D-26)\, e^{\fft12 a\, \hat \phi}\, 
{\hat *\oneone}\,.
\label{dlagc}
\ee
We shall refer to this extra contribution as a ``cosmological term.''
Note that if we were instead considering the theory of gravity, 3-form
and dilaton as coming from the low-energy effective theory of the
superstring, the $(D-26)$ factor would be replaced by $(D-10)$.  In
all subsequent discussions in this section, 26 can accordingly be
replaced by 10 in the context of the superstring.

   It is of interest to see what happens to the previous Kaluza-Klein
reductions on $S^3$ and $S^{D-3}$ after this extra term is included.
We find that the previous $S^3$ reduction Ansatz continues to give a
consistent reduction, in which all the dependence on the $S^3$
coordinates cancels out when it is substituted into the
$D$-dimensional equations of motion following from (\ref{dlagc}).  We
find that the reduced $(D-3)$-dimensional theory is described by the
same Lagrangian (\ref{dm3lag}), but now the potential $V$ given in
(\ref{s3pot}) is replaced by
\be V = \ft12 g^2\, Y^{\fft12}\, \Big(2
\wtd T_{ij}\, \wtd T_{ij} - (\wtd T_{ii})^2 \Big) + \ft12 m^2\,
(D-26)\, Y^{\fft12} \,.\label{s3potc} 
\ee 

    The fact that the $S^3$ reduction continues to be a consistent one
after the inclusion of the cosmological term in (\ref{dlagc}) could in
fact have been foreseen by considering the group-theory arguments that
we developed in section 2.  In the absence of the cosmological term,
we observed that the global symmetry group after a $T^3$ reduction is
$\R\times O(3,3)$, which is large enough to contain $O(3)\times O(3)$
as a compact subgroup, and hence to permit an $SO(4)$ gauging.  The
inclusion of the cosmological term in (\ref{dlagc}) breaks the $\R$
factor in the global symmetry, but the $O(3,3)$ factor
survives,\footnote{This can be seen from the fact that the dilaton
vector for the cosmological term after the $T^3$ reduction is
orthogonal to the dilaton vectors that form the positive roots of
$O(3,3)$.} and so the cosmological term does not present any obstacle
to the $SO(4)$ gauging in $D-3$ dimensions.

  It is interesting to note that if $D>26$ (or $D>10$ in the case of a
supersymmetric string), the potential
(\ref{s3potc}) admits a symmetrical ground-state solution in which all
the scalar fields are constant.  To see this, we note that for such a
solution we must have
\be
\fft{\del V}{\del Y} = 0\,,\qquad \fft{\del V}{\del \wtd T_{ij}} -
\ft14 \delta_{ij} \, \delta_{k\ell}\,  \fft{\del V}{\del \wtd T_{k\ell}}
=0\,.
\ee
(The trace subtraction in the second equation arises because $\wtd
T_{ij}$ has unit determinant.)  Thus the conditions for a solution
with constant scalars imply
\be
V= 0\,, \qquad \wtd T_{ij} = \ft14 \wtd T_{kk}\, \delta_{ij}\,,
\ee
and hence since $\wtd T_{ij}$ is unimodular we must have $\wtd
T_{ij}=\delta_{ij}$, and 
\be
g= m\, \sqrt{\fft{D-26}{8}}\,,
\ee
with $Y$ arbitrary.  Note in particular that the vanishing of $V$
implies that the $(D-3)$-dimensional Einstein equation has no
cosmological term, and so it admits Minkowski spacetime as a
ground-state solution.  One can also find non-trivial solutions that
are asymptotically flat.

    If we now consider instead the $S^{D-3}$ reduction of the new
theory (\ref{dlagc}), we find that the previously consistent reduction
is spoiled by the presence of the additional cosmological term.  In
particular, it turns out that there is a mis-match between the
$S^{D-3}$ dependence from the extra term $e^{\fft12 a\, \hat \phi}\,
{\hat *\oneone}$, in comparison to the previous terms, in the
$D$-dimensional equation of motion for the dilaton $\hat\phi$.
Actually, this is not too surprising.  It can be understood from the
fact that the presence of the cosmological term in (\ref{dlagc})
breaks the enhanced $O(D-2,D-2)$ global symmetry that occurred
previously under a dimensional reduction on $T^{D-3}$, and so there
will no longer be an $SO(D-2)$ compact subgroup of the global symmetry
group that could permit an $SO(D-2)$ gauging in three dimensions.
This can be seen from the fact that the dilaton vector for the
cosmological term in (\ref{dlagc}), after toroidal reduction on
$T^{D-3}$, is not orthogonal to the positive root vectors of
$O(D-2,D-2)$.  

   Finally, we may also consider the possible inclusion of an
analogous cosmological term in the Lagrangian (\ref{dlagf2}) for
gravity, the dilaton and a 2-form field strength.  In this case there
would not be any direct motivation from bosonic string theory for the
inclusion of such a term, but it is nevertheless of interest to see
what the effect would be.  Thus we may
consider whether we may modify the Lagrangian (\ref{dlagf2}) to
\be
{\cal L}_D = \hat R\, {\hat *\oneone} - 
\ft12 {\hat *d\hat\phi}\wedge d\hat\phi  
-\ft12 e^{-a\, \hat\phi}\, {\hat *\hat F_\2}\wedge \hat F_\2
-\ft12 m^2\, e^{b\, \hat\phi}\, {\hat *\oneone}\,,
\label{dlagf2c}
\ee
where the dilaton coupling constant $b$ in the cosmological term is
chosen so as to maintain the consistency of the Kaluza-Klein $S^2$
reduction.  It turns out that this is indeed possible, and 
consistency is achieved if $b$ is the positive constant given by
\be
b^2 = \fft{2}{(D-1)(D-2)}\,.\label{bvalue}
\ee
The resulting Kaluza-Klein theory in $(D-2)$ dimensions is described
by the Lagrangian (\ref{dm2lag}), but with the potential
(\ref{dm2pot}) replaced by 
\be
V = \ft12 g^2\, Y^{\fft23}\, \Big(2 \wtd T_{ij}\, \wtd T_{ij} 
  - (\wtd T_{ii})^2 \Big) + \ft12 m^2\, Y^{\fft2{D-2}}
\,.\label{dm2potc}
\ee

    Again, one could have foreseen the continued consistency of the $S^2$
reduction from the fact that if the theory (\ref{dlagf2c}) is reduced
instead on $T^2$, there is still a sufficient enhancement of the
global symmetry to permit an $SO(3)$ gauging.  Previously, for
(\ref{dlagf2}), the generic $GL(2,\R)$ symmetry was enhanced to to
$GL(3,\R)$.  Now, with the inclusion of the cosmological term in
(\ref{dlagf2c}), the $\R$ factor in the $GL(3,\R)$ is broken, but the
$SL(3,\R)$ factor remains, and so the compact $SO(3)$ subgroup is
still available for the gauging.  We can also understand this as
follows.  Recalling that the original Lagrangian (\ref{dlagf2}) can
itself be viewed as a standard $S^1$ reduction of pure gravity in
$(D+1)$ dimensions, we now observe that the enlarged Lagrangian
(\ref{dlagf2c}), with $b$ given by (\ref{bvalue}), is nothing but the
$S^1$ reduction of the $(D+1)$-dimensional theory of pure gravity with
a cosmological constant:
\be
{\cal L}_{D+1} = \hat R_{D+1}\, {\hat *\oneone} - \ft12 m^2\, {\hat
*\oneone}\,.\label{d1lagc}
\ee
It is then evident that the dimensional reduction of (\ref{dlagf2c})
on $T^2$ will give the same theory as the dimensional reduction of
(\ref{d1lagc}) on $T^3$, and so in particular there will be a
$SL(3,\R)$ global symmetry.\footnote{The cosmological constant in
$(D+1)$ dimensions breaks the scale-covariance that a theory of
gravity and antisymmetric tensors has, and so one only gets
$SL(n,\R)$, and not $GL(n,\R)$ from a $T^n$ reduction in this case
(see \cite{cjlp1}).}

    One can again look for solutions of the reduced theory in which
all the scalars are constant.  The equations of motion following from
(\ref{dm2potc}) then imply that
\be
\wtd T_{ij}= \delta_{ij}\,,\qquad Y^{\fft{2(D-5)}{D-2}} = \fft{m^2}{g^2\,
(D-2)}\,.\label{tysol}
\ee
Substituting these back into the potential, we find that at this
extremum it is give by
\be
V= \ft12 m^2\, \Big(\fft{D-5}{D-2}\Big)\, \Big[ \fft{m^2}{g^2\,
(D-5)}\Big]^{\fft1{D-5}}\,,
\ee
which corresponds (for $D\ge 6$) to a positive cosmological constant
in the $(D-2)$-dimensional spacetime.  (Note that (\ref{tysol})
implies that the cosmological constant in the $(D+1)$-dimensional pure
gravity theory is also positive.)  This allows, in particular, a
ground-state solution of the original $D$-dimensional theory of the
form $M_{D-2}\times S^2$, where $M_{D-2}$ is an Einstein spacetime
with positive cosmological constant, such as de Sitter space.
Interpreted as a solution of the $(D+1)$-dimensional pure Einstein
theory with positive cosmological constant, it becomes $M_{D-2}\times
S^3$, since in this solution the 2-form $\hat F_\2$ in $D$ dimensions
is a constant multiple of the volume-form of $S^2$, and thus the $S^1$
in the reduction from $(D+1)$ dimensions is the Hopf bundle over $S^2$.

\section{Conclusions and discussions}

     In this paper, we have investigated the consistency of the
Kaluza-Klein sphere reduction of the theory described by the
Lagrangian (\ref{dilatonlag}), comprising gravity coupled to a
$p$-form field strength and a dilaton in $D$ dimensions.
Specifically, we have focussed our attention on those cases where the
reduction Ansatz at least includes all the Yang-Mills fields of the $SO(n+1)$
gauge group.

    We have shown that by including the dilaton in the
higher-dimensional theory, the possibilities for consistent sphere
reductions are extended somewhat, in comparison to the case where the
higher-dimensional starting point comprises only gravity and a
$p$-form field strength.  Specifically, if no dilaton is included the
only possibilities for consistent sphere reductions of the kind we are
considering are those associated with the $S^4$ and $S^7$ reductions
of $D=11$ supergravity, and the $S^5$ reduction of type IIB
supergravity.  With the dilaton included, we find that consistent
$S^2$ reductions are possible for the case of a 2-form in the
higher dimension $D$, and that consistent $S^3$ and $S^{D-3}$
reductions are possible for the case of a 3-form in the
higher-dimension.  These reductions are possible starting from an
arbitrary dimension $D$, provided that the strength of the dilaton
coupling to the 2-form or 3-form field strength is chosen
appropriately.  

   The previously-known consistent sphere reductions from $D=11$ with
a 4-form, and $D=10$ with a self-dual 5-form, were associated with
supersymmetric higher-dimensional theories.  In the examples that we
have obtained in this paper, supersymmetry is clearly not in general
playing a r\^ole, since the higher-dimensional starting point can be a
theory of gravity, a dilaton and a 2-form or 3-form in any arbitrary
dimension.  It is probably more appropriate, therefore, to
characterise the theories that admit consistent sphere reductions by
the fact that they have the unusual property of giving rise to
lower-dimensional theories with certain enhanced global symmetry
groups upon toroidal reduction on $T^n$.  In particular, a necessary
condition for a consistent $n$-sphere reduction that retains all the
Yang-Mills fields of $SO(n+1)$ is that the global symmetry $GL(n,\R)$
of a generic theory reduced on $T^n$ must be enhanced to a group whose
compact subgroup contains $SO(n+1)$.  These symmetry enhancements
occur only in exceptional cases, when scalars coming from the toroidal
reduction of metric ``conspire'' with scalars coming from the
reduction of the $p$-form field strength to give an enhanced global
symmetry group.  It so happens that this same feature of symmetry
enhancement is a central feature also in theories such as $D=11$ and
type IIB supergravity, and their toroidal reductions.

   It should be emphasised that the group-theoretic argument that we
have been using in order to determine when a consistent sphere
reduction may be possible does not, of itself, provide a guarantee of
consistency.\footnote{Unlike the traditional group theory argument
that proves conclusively the consistency of a truncation in which all
singlets under a symmetry group are retained, and all non-singlets are
truncated.}  Rather, it can be viewed as providing a proof of {\it
inconsistency} in cases where the necessary enhancement of the global
symmetry group in the associated toroidal reduction does not occur.
It is rather striking, however, that in all cases where a suitable
sufficiently large global symmetry enhancement does occur, we find
that a consistent sphere reduction is possible.

    An interesting illustration of this point is provided by the
reductions that we considered in section 7, where an additional
``cosmological term'' was included in the higher-dimensional theory.
The argument based on global symmetry enhancement showed that the
$S^{D-3}$ reduction would no longer be consistent, but that the $S^3$
and $S^2$ reductions still had the possibility of being consistent.
And indeed this is just what we found, when we substituted the
Ans\"atze into the equations of motion of the higher-dimensional
theories with the cosmological terms included.

   Although we have argued that supersymmetry is in some sense not
of itself the directly crucial ingredient in the question of consistency,
it is, nevertheless, worthwhile to consider further the question of
supersymmetry and consistent sphere reductions.  As well as the
examples of the $S^4$ and $S^7$ reductions from $D=11$, and the $S^5$
reduction from $D=10$, we can now also consider those examples amongst
the reductions constructed in this paper that can be associated with
supersymmetric theories.  Thus, for instance, we can consider the
$S^2$ reduction of type IIA supergravity, using the R-R 2-form, and
the $S^3$ and $S^7$ reductions of type I or type II supergravity,
using the NS-NS (or R-R in the case of type IIB) 3-form.

    Constructing the Kaluza-Klein sphere reduction Ansatz for the
fermions in a supergravity theory is a notoriously difficult problem,
and even when it is attempted the efforts are rarely extended to
include the quartic fermion terms.  However, we may construct a
general argument to show that once a consistent reduction has been
constructed in the bosonic sector, the supersymmetry of the
higher-dimensional theory will then guarantee that a consistent
reduction including the fermions as well must be possible.  The
argument is as follows.  We know that a sphere reduction in which {\it
all} fields (massive as well as massless) are retained will
necessarily be consistent, and it will give rise to a supersymmetric
lower-dimensional theory.   Furthermore, we know that all the
non-linear couplings between the various lower-dimensional fields will
be organised, by virtue of the lower-dimensional supersymmetry, into
supersymmetrically-covariant couplings of complete supermultiplets.
Now, if we demonstrate in the bosonic sector that there is a
consistent truncation to the massless sector (\ie to the bosonic
sector of the massless lower-dimensional supermultiplet), then this
means that there are no interaction terms in which powers of the massless
bosonic fields (\ie conserved currents built from the massless fields
(see \cite{zilch})) couple to linear powers of the massive bosons that are
being set to zero.  But this in turn implies that in the full theory
there can be no interaction terms in which supercurrents built from
the massless multiplet couple to linear powers of the massive fields.
Thus if one shows that it is consistent to make a sphere reduction in
which all the bosons of the massless supermultiplet are retained, then
this implies that it must be consistent to make a sphere reduction of
the supersymmetric theory in which the entire massless supermultiplet
is retained.

    One can use this argument to show that the $S^3$ and $S^7$
reductions of $N=1$ ten-dimensional supergravity, which are special
cases of our more general results in this paper, will be consistent,
as a consequence of our results for the bosonic sectors.

\section*{Acknowledgement}  C.N.P. is grateful to the University of
Pennsylvania for hospitality during the course of this work.  We are
grateful to Arta Sadrzadeh and Tuan Tran for extensive discussions.

\end{document}